\documentclass{article}

\usepackage{arxiv}

\usepackage[utf8]{inputenc} 
\usepackage[T1]{fontenc}    
\usepackage{hyperref}       
\usepackage{url}            
\usepackage{booktabs}       
\usepackage{amsfonts}       
\usepackage{nicefrac}       
\usepackage{microtype}      
\usepackage{graphicx}
\usepackage{doi}

\title{Advancing Global South University Education\\ with Large Language Models}

\date{} 					

\author{
{\hspace{1mm}Kemas Muslim L}
        \thanks{Corresponding author email: kemasmuslim@telkomuniversity.ac.id} \\
	School of Computing\\
	Telkom University\\
	Bandung, Indonesia \\
	\And
{\hspace{1mm}Toru Ishida} \\
	School of Computing\\
	Telkom University\\
	Bandung, Indonesia \\
	\And
{\hspace{1mm}Aditya Firman Ihsan} \\
	School of Computing\\
	Telkom University\\
	Bandung, Indonesia \\
	\And
{\hspace{1mm}Rikman Aherliwan Rudawan} \\
	School of Applied Sciences\\
	Telkom University\\
	Bandung, Indonesia \\
}

\renewcommand{\headeright}{}
\renewcommand{\undertitle}{}
\renewcommand{\shorttitle}{Advancing Global South University Education with Large Language Models}

\hypersetup{
pdftitle={Advancing Global South University Education with Large Language Models},
pdfauthor={Kemas Muslim L, Toru Ishida, Aditya Firman Ihsan, Rikman Aherliwan Rudawan},
pdfkeywords={the Global South, higher education, large language models, ChatGPT},
}

\begin{document}
\maketitle

\begin{abstract}
In recent years, it has been observed that the center of gravity for the volume of higher education has shifted to the Global South. However, research indicates a widening disparity in the quality of higher education between the Global South and the Global North. Although investments in higher education within the Global South have increased, the rapid surge in student numbers has resulted in a decline in public expenditure per student. For instance, the student-to-teacher ratio in the Global South is significantly higher compared to that in the Global North, which poses a substantial barrier to the implementation of creative education. In response, Telkom University in Indonesia has embarked on an experiment to enhance the quality of learning and teaching by integrating large language models (LLMs) such as ChatGPT into five of its courses—Mathematics, English, Computing, Computer Systems, and Creative Media. This article elucidates the ongoing experimental plan and explores how the integration of LLMs could contribute to addressing the challenges currently faced by higher education in the Global South. 
\end{abstract}


\section{Introduction}

Although the term "Global South" does not have a universally accepted definition, it is generally used to refer to developing countries~\cite{kowalski2020globalsouth}. Despite the various challenges faced by these nations, the Global South encompasses a large number of countries and populations, surpassing those of the Global North in scale. Therefore, the development of this region is crucial not only for the countries within it but also for the stability, sustainability, prosperity, and well-being of the entire world. In particular, quality education is essential for the Global South to achieve sustainable development~\cite{uvalic2016sustainable}. In this article, we will first survey the issues faced by higher education in the Global South.

\section{ISSUES IN HIGHER EDUCATION IN THE GLOBAL SOUTH}
The Global South comprises nations with diverse characteristics, yet they face the following common challenges that have been widely recognized.

\subsection{Widening Gap in Higher Education Quality Between the Global South and the Global North}

\begin{figure}
\centering
\includegraphics[scale=0.9]{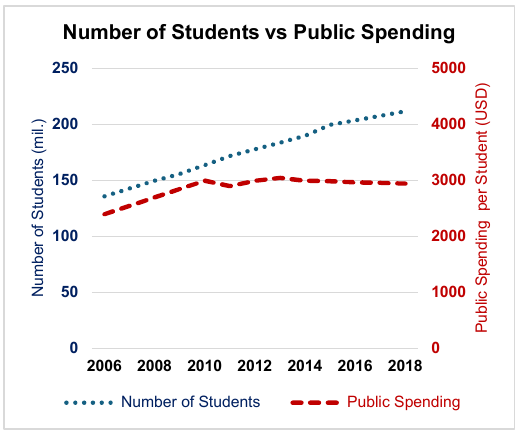}
\caption[]{A comparison of the total number of students (in millions) and public spending per student (in USD) in the Global South. While the student population continues to grow, public spending per student has not kept pace, resulting in a widening gap between the two. The data in this graph is based on the 2022 HESA report on global higher education institutions, students, and funding~\cite{hesa2022world}.}
\label{fig:nbofstudvsps}
\end{figure}

Over the past two decades, the disparity in the quality of higher education between the Global South and the Global North has been steadily increasing. While the number of students in the Global South continues to rise, public spending on higher education has remained relatively stagnant (Figure~\ref{fig:nbofstudvsps}). In contrast, the Global North has seen consistent increases in public expenditure on higher education, despite stable student numbers. In other words, while funding per student in the Global South is declining, it is increasing in the Global North, thereby exacerbating the quality gap between the two regions. This widening disparity has been highlighted in a report by Higher Education Strategy Associates (HESA)~\cite{hesa2022world}, as noted in an interview with Alex Usher, the president of the investigating body~\cite{macgregor2022higher}.

\subsection{High Student-to-Faculty Ratios in the Global South Higher Education Institutions Leading to Overburdened Faculty and Insufficient Student Engagement}

The massification of higher education over the past 50 years has led to an increase in student-to-faculty ratios worldwide, a trend particularly pronounced in the Global South~\cite{buckner2021quantity}. Although this metric does not directly reflect the quality of higher education institutions, it is commonly used by governments and international higher education rankings to assess the capacity of national education systems~\cite{schenker2012economics}. When the student-to-faculty ratio is high, as it is in many Global South countries~\cite{buckner2021quantity}, faculty members face an increased administrative burden in addition to their core teaching and research responsibilities. This, coupled with large student numbers, reduces the level of engagement between faculty and students, thereby diminishing the quality of education~\cite{schenker2012economics}.

To address these challenges, it is imperative to explore innovative solutions. Recently, generative AI tools have emerged, predicted to have a disruptive impact on education. While the application of generative AI in education has sparked ethical debates, there is potential for these tools to mitigate some of the challenges faced by the Global South. On the other hand, initiatives to integrate AI into education in the Global South are very limited, with only a few countries making notable progress in AI adoption~\cite{minaee2024largelanguage}. Through pilot projects that actively integrate large language models (LLMs) into various university courses, we aim to enhance the quality of learning in higher education across the Global South.

\section{PROS AND CONS OF INTEGRATING LLMS IN EDUCATION}

AI has been utilized in education long before the emergence of generative AI. Among the various branches of AI, generative AI has gained significant traction, especially since the advent of publicly accessible LLM applications such as ChatGPT~\cite{minaee2024largelanguage}. In higher education, AI tools are employed both at the core of educational activities, such as teaching and learning, and in supporting roles, such as educational administration~\cite{liu2023harnessing, kshetri2023future}. Due to its benefits and opportunities, there has been a surge in research on the use of LLMs in education. LLMs have been applied in various courses, including language learning~\cite{li2024selfdirected}, science~\cite{lee2024chatgpt}, multimedia~\cite{chen2024multimedia}, and computer science~\cite{rodriguez2024chatgpt}, and are also being used to enhance teaching and learning methods~\cite{wang2023unleashing}.

Despite their promising potential and the opportunity to integrate them into educational activities, the use of LLMs in education presents both advantages and challenges. These factors should be carefully considered by educational institutions, learners, and educators aiming to leverage LLMs to improve educational quality.

The advantages of using LLMs in education include:
\begin{enumerate}
\item Interactive learning~\cite{ahmad2023generative, mosaiyebzadeh2023chatgpt}\\
As conversational AI tools, LLMs offer an interactive learning experience, allowing students to ask questions and receive instructions for completing tasks that support their assignments.
\item Personalized content and interaction~\cite{ahmad2023generative, mosaiyebzadeh2023chatgpt}\\
LLMs facilitate personalized interactions, enabling students to learn at their own pace and tailor content to meet their specific needs.
\item Reducing educators' workload~\cite{mosaiyebzadeh2023chatgpt}\\
LLMs can function as learning assistants, reducing the workload of educators. Additionally, LLMs and other AI tools support various educational tasks that may not be directly related to teaching, such as designing lesson plans, preparing teaching materials, generating question banks, grading, and more.
\item Multilingual support~\cite{ahmad2023generative, mosaiyebzadeh2023chatgpt}\\
LLMs offer multilingual capabilities, allowing for conversations in multiple languages and translation between languages. This feature is particularly beneficial for students in the Global South, where English may not be the primary language.
\end{enumerate}

However, despite the enthusiasm for embracing LLMs, several challenges limit their full potential:

\begin{enumerate}
     
\item Ethical Issues~\cite{ahmad2023generative}\\
The content generation capabilities of LLMs can be misused by both students and educators, making it difficult to assess the originality of students' work. This has led some educators to either prohibit the use of LLMs entirely or design assignments and grading methods that are compatible with LLM usage.
\item Data Privacy and Security~\cite{mosaiyebzadeh2023chatgpt}\\
Concerns about data privacy and security arise from the use of sensitive personal information and the measures needed to protect such data. Therefore, adopters of LLMs must implement appropriate safeguards to ensure data management complies with the regulations of their respective countries.
\item Reliability and Accuracy~\cite{ahmad2023generative, mosaiyebzadeh2023chatgpt}\\
The current state of LLMs includes issues with reliability and accuracy, such as hallucinations and biases. Consequently, both students and educators should verify the accuracy of the generated responses.
\item Cost\\
Although many prominent LLMs offer limited free features, more advanced versions often require a subscription. In the Global South, particularly in low-income countries, universities, students, and educators may be hesitant to subscribe to paid versions of LLM applications.
\end{enumerate}

\begin{figure}
\centering
\includegraphics[scale=0.7]{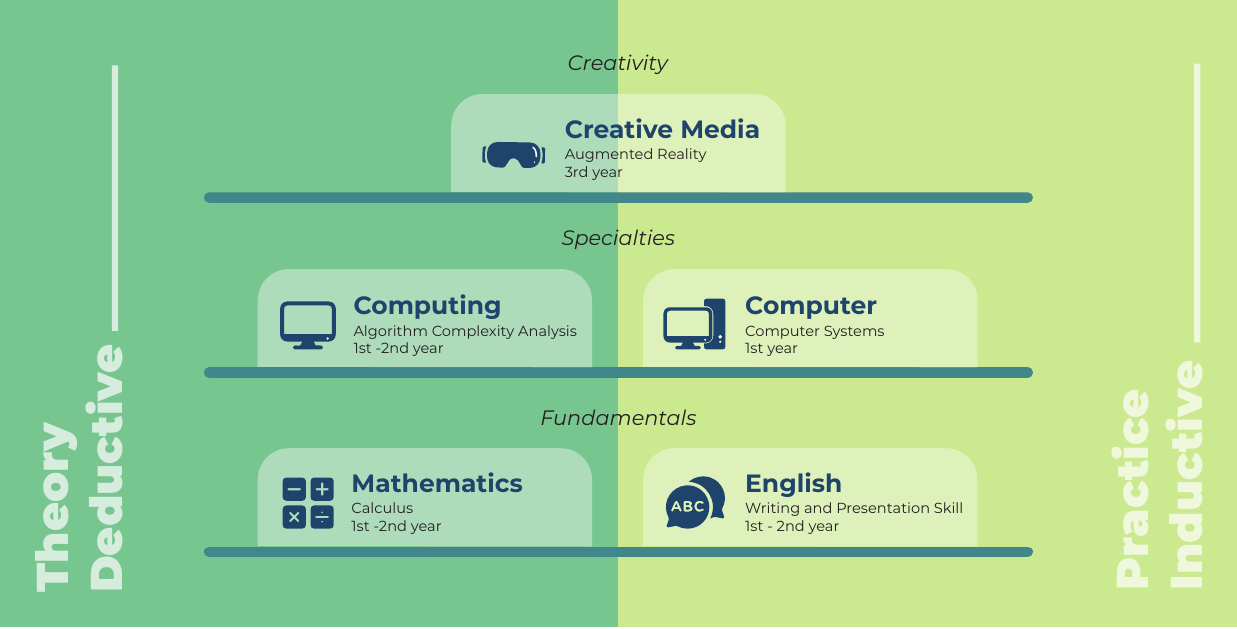}
\caption[]{Five courses have been selected for the pilot study to integrate LLM applications into the learning process. These courses represent a typical computer science curriculum, encompassing both theory/deductive and practice/inductive approaches. Vertically, these courses span different levels: foundational courses (Mathematics and English), specialized courses (Computing and Computer Systems), and creative courses (Creative Media).}
\label{fig:courses}
\end{figure}

\section{REAL SCALE EXPERIMENTS IN INDONESIAN TELKOM UNIVERSITY}

The application of generative AI in education has been explored in various contexts, but in the Global South, reducing the burden on educators is a critical condition for its successful adoption. Implementing generative AI in a manner that increases the workload for educators would likely be unacceptable in educational settings. However, it is equally important that the introduction of generative AI contributes to improving the quality of education, as its deployment would be counterproductive if it leads to a decline in educational standards. While generative AI may directly enhance student outcomes and motivation, it could also have indirect benefits, such as increasing the time educators have to engage more deeply with students by making educational processes more efficient.

Given this context, rather than merely replicating research conducted in the Global North, we have chosen to start with the challenges faced by our project members at Telkom University in Indonesia and generalize these challenges into research topics. Five courses from three faculties at Telkom University were selected as subjects for the pilot study, representing a diverse range of curricula within the university.

Figure~\ref{fig:courses} illustrates these five courses and the categories of curricula they represent. The Mathematics and English courses are general education courses required for first-year students. The Computing and Computer courses focus on specialized technical skills, while the Creative Media course applies these skills to create societal impact.

We will conduct the pilot project during the actual academic term, collecting and analyzing data to clarify the effectiveness of using generative AI. Specifically, we will carry out action research by utilizing two classes for each of these five courses: one will serve as the experimental class, and the other as the control class.

The research has two primary objectives. The first is to design interactions between students, LLMs, and educators, and to establish methodologies for learning support using LLMs. The second objective is to obtain empirical evidence that the introduction of LLMs in higher education can both alleviate instructors' workload and enhance student learning. Our hypothesis is that when engagement between students, LLMs, and instructors is appropriately designed: a) the burden on educators will be reduced, and b) students' learning motivation will improve. To test this hypothesis, we will set up the following experimental environment:

\subsection{Experimental Environment}

To test this hypothesis, we will set up the following experimental environment:
\begin{enumerate}
    \item \textbf{Courses and classes:} Five courses where LLM will be applied are selected for the pilot study. Each course will have two concurrently conducted classes: one integrating LLM (experimental class) and the other without LLM usage (control class), making a total of ten classes for the experiment.
    \item \textbf{Learning assistant application:} A student-LLM-instructor interface application will be developed for this experiment. This interactive application aims to enhance educational efficiency while improving the quality of student learning.
    \item \textbf{Pre-prepared prompts:} Instructors will prepare prompts for students to start interacting with the LLM through the interface. Students will execute these prompts via the interface, and instructors will monitor the interactions between students and the LLM. The interface application will record the interactions among students, LLM, and instructors to aid in future improvements of teaching methodologies.
\end{enumerate}

\begin{figure}
\centering
\includegraphics[scale=0.7]{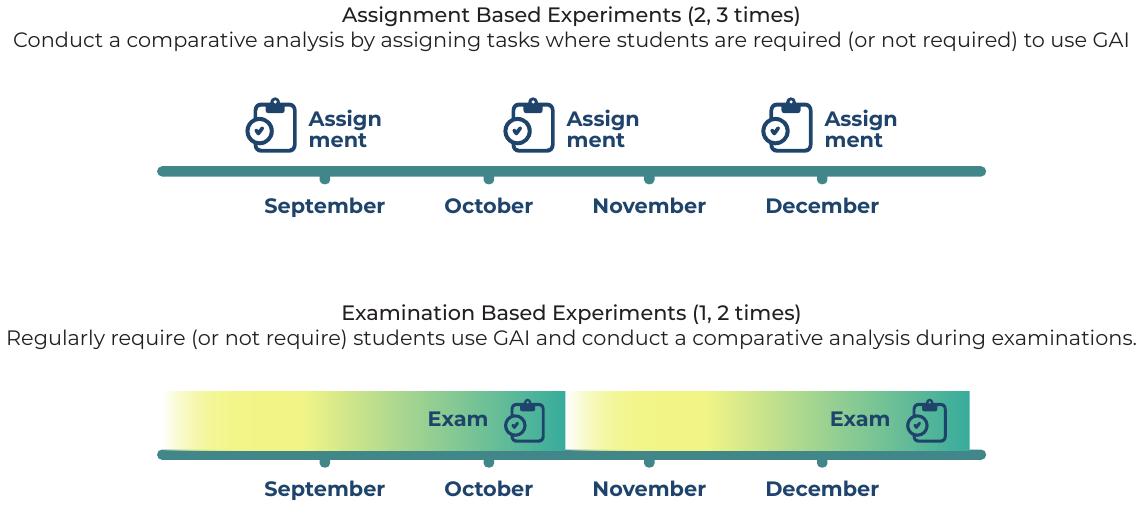}
\caption[]{Two types of experiments: assignment-based and examination-based. The first utilizes assignments to evaluate students achievement, whereas the latter conduct examinations. Both methods are applied in the experimental and control classes.}
\label{fig:experiment}
\end{figure}

\subsection{Two Types of Experiment}

The experiments for the five courses will be conducted as shown in Figure~\ref{fig:experiment}. There are two types of experiments:
\begin{enumerate}
    \item \textbf{Assignment-Based Experiment:} This experiment will be conducted using assignments given during the course. The same assignments will be given to both the experimental class and the control class, and students will complete them under the above two conditions. The results will then be evaluated.
    \item \textbf{Examination-Based Experiment:} The course will be divided into several periods, and experiments will be conducted during each period, with evaluations made using exams at the end of each period. Simultaneously, the Motivated Strategies for Learning Questionnaire (MSLQ)~\cite{pintrich1988motivated} will be administered to compare student motivation at the end of each period under the two conditions.
\end{enumerate}

\subsection{Experiment Sessions}

To effectively integrate LLM as a learning assistant, each experimental class will be structured into three distinct sessions:
\begin{enumerate}
    \item \textbf{Introduction Session:} Instructors will provide an overview of the lecture, outline the expected Course Learning Outcomes (CLOs), and present the lecture agenda.
    \item \textbf{LLM-Assisted Session:} Students will be assigned tasks to complete during class with the assistance of an LLM. These tasks may include self-directed learning activities, solving problem sets, working on exercises, or summarizing learning materials. Pre-prepared prompts will be provided to guide students in their interactions with the LLM. During this session, instructors will monitor the exchanges between students and the LLM to ensure that the prompts and responses are relevant and contribute effectively to the CLOs. Students will interact with the LLM through an application that facilitates the interaction between students, the LLM, and educators, which will be discussed next.
    \item \textbf{Engagement Session:} This session will involve traditional interaction between students and instructors to assess the outcomes of the LLM-assisted session and ensure that the CLOs are being met. This interaction will serve to evaluate the previous session and confirm that the learning objectives have been achieved.
\end{enumerate}

\section{CLASSIFICATION OF COURSES AND THEIR ISSUES IN THE GLOBAL SOUTH CONTEXT}

Following the subject classification shown in Figure~\ref{fig:courses}, this section addresses the issues targeted by applying LLMs, focusing on challenges in the Global South. First, we address the issues related to the courses according to the horizontal classification, i.e. theory/deductive and practice/inductive approaches:

\textbf{Theory/Deductive:} In subjects categorized under this classification, such as mathematics, understanding abstract and formal concepts is critical. For instance, if students do not fully grasp the concept of "differentiation," no amount of practice solving differentiation problems will enable them to tackle new problems effectively. The same applies in computing: unless students understand the concept of "recursion" in relation to recurrence relations, they will struggle to develop programs using this concept. To help students comprehend these abstract and formal concepts, they need a variety of explanations, including metaphors, analogies, or concrete examples, from which they can find something that resonates with them. However, in universities in the Global South, teachers often handle large classes and cannot provide detailed explanations or engage in extensive question-and-answer sessions with individual students. By generating various explanations and responding to students' questions in place of the teacher, LLMs have the potential to promote a deeper understanding of abstract and formal concepts.

\textbf{Practice/Inductive:} In this category, subjects such as English require students to learn a wide range of examples in order to grasp higher-level concepts and operational rules, despite the presence of exceptions. For instance, students learn the concept and grammar of the "present perfect tense" through example sentences. However, simply memorizing grammar rules is insufficient for understanding or composing English in practice. The same occurs in computing. When building a computer system, it is essential to know the functions and operating methods of its components. However, memorizing the vast and diverse functions and methods defined for these devices is possible only for a small number of experts. Most engineers must refer to extensive manuals to understand the necessary information at each step, and their ability to comprehend these manuals depends on their grasp of higher-level concepts such as architecture and design philosophy. In universities in the Global South, teachers are often unable to provide students with selected information from manuals. By selecting and providing key information, along with higher-level concepts, LLMs can enhance students’ understanding and bridge this gap.

\begin{figure}
\centering
\includegraphics[scale=0.75]{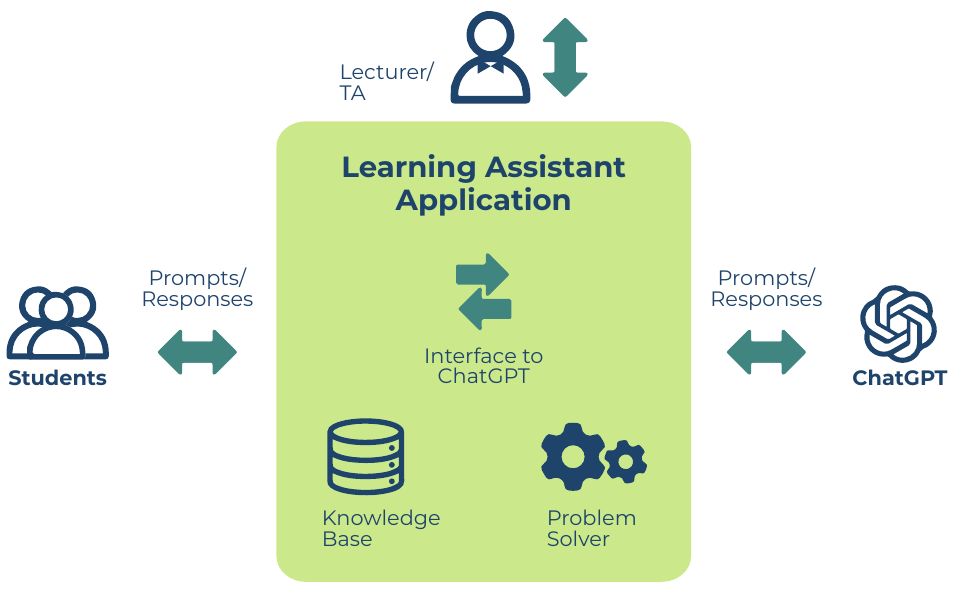}
\caption[]{A learning assistant application is needed to facilitate the tripartite interaction between students, LLMs, and educators. This allows educators, including lecturers and teaching assistants, to monitor student-LLM interactions and intervene when necessary. The application also features a knowledge base for recording course-specific information, a question bank, and an assignment repository. Additionally, the problem solver module helps guide students through step-by-step activities required to solve complex problems.}
\label{fig:application}
\end{figure}

From the perspective of fundamentals, specialties, and creativity subjects, the issues in the courses are discussed as follows:

\textbf{Fundamentals:} In technical curricula, foundational subjects such as mathematics and English serve as critical training for acquiring specialized knowledge. If students perform poorly in these subjects, they will struggle with more advanced, specialized courses. Therefore, supporting students with lower grades is especially important, but in universities in the Global South, teachers often lack the time and resources to provide this support. By offering repeated exercises tailored to students' skill levels and assisting with follow-up support, LLMs can significantly enhance the quality of learning for students and improve the efficiency of teaching for instructors.

\textbf{Specialties:} Specialized courses form the core of technical curricula, and they are typically offered as a combination of lectures and practical exercises. These exercises may be conducted individually or in teams. Ideally, each student should complete the exercises independently, as everyone needs to develop their own expertise. However, monitoring and providing individual instruction during exercises is a heavy burden for teachers, particularly in universities in the Global South. As a result, team-based exercises, which allow for knowledge sharing among students, are often implemented. However, if knowledge sharing is not conducted effectively, the benefit of these exercises diminishes. By using LLMs to shift from team-based to individual exercises, it becomes possible to strengthen students’ expertise in a more targeted manner.

\textbf{Creativity:} In technical curricula, courses are often designed to encourage students to apply their specialized skills to develop new products and services. These courses incorporate project-based learning, where group work fosters student creativity. In the Global North, interdisciplinary teams of instructors guide students, and teaching assistants (TAs) are assigned as facilitators for each group, with increasing emphasis on this collaborative support. In contrast, in the Global South, although the importance of group work is recognized, there are often insufficient teachers and TAs to support the process. As a result, a few students tend to lead the group while others take a passive role, which reduces overall effectiveness. LLMs can facilitate student collaboration, improving the quality of group work and greatly enhancing educational outcomes.

While the challenges of introducing LLMs vary across categories, each subject area may have its own specific concerns. For example, in mathematics, many educators express doubt as to whether AI can truly improve students' problem-solving abilities. While it may help students become more proficient at calculations, some fear that it could negatively impact their ability to solve core problems. In other words, there is concern that students who effectively use ChatGPT might develop procedural skills at the expense of their problem-solving abilities. In contrast, in English, the shocking reality is that many students already generate most of their assignments using LLMs. Consequently, training based on outcome-oriented evaluation is no longer viable, and immediate changes are required. With the significance of writing assignments fading, a shift from outcome-based education to process-oriented learning is urgently needed. In creative media, students annually modify and extend examples from cookbooks to develop new products and services. However, this often leads to errors, as students apply examples without questioning them. While it may be unrealistic to expect expert knowledge to be transmitted through examples alone, LLMs can take on the role of educators, using their knowledge and dialogue capabilities to help students better understand scripts and improve their creative output.

\section{APPLICATION INTERFACE FOR THE CLASS ACTION RESEARCH ENVIRONMENT}

To effectively integrate LLM into the classroom as learning assistants, an application is needed to facilitate the tripartite interaction between students, LLM, and educators (Figure~\ref{fig:application}). This application serves as an interface to address the following challenges that arise when students interact with LLMs:
\begin{enumerate}
    \item Varying levels of prompt engineering skills among students.\\
    In an LLM-supported learning environment, students need both prerequisite knowledge and prompt engineering skills~\cite{wang2023unleashing}. Well-constructed prompts increase the likelihood of receiving high-quality responses, while poorly crafted prompts tend to yield low-quality results. Since students possess different levels of these skills, educators must guide them in formulating effective prompts during in-class activities~\cite{wang2023unleashing}.
    \item Inappropriate prompts and responses.\\
    In the Global South higher education institutions, where class sizes often exceed 40 students, it is quite challenging for educators to ensure that all students provide appropriate prompts when using LLMs. Additionally, LLMs may return inaccurate responses, requiring instructors to correct and provide accurate information to students.
\end{enumerate}

To address these challenges, the application should offer essential features to facilitate the tripartite interaction, as shown in Figure~\ref{fig:application}. The figure presents the interaction model between students, educators, and a learning assistance application that bridges their interaction with ChatGPT. The application communicates directly with ChatGPT, leveraging its capabilities to assist both students and educators. On the top side, an educator oversees the interaction, guiding students in using the application effectively. ChatGPT provides responses based on the prompts entered, and educators can intervene to ensure appropriate guidance and correction when necessary. This setup allows for a structured, collaborative learning experience where students benefit from both AI and educator oversight.

The main features of the application are following:
\begin{enumerate}
    \item \textbf{Interface to ChatGPT:} During class activities, students are only allowed to give prompts to ChatGPT through this application. This allows educators to monitor all the conversations between the students and ChatGPT in realtime, and to intervene with the interaction when necessary. At the left of Figure~\ref{fig:application}, students are depicted engaging with the system to give prompts and getting responses from ChatGPT via the application, which is the central element of the application that bridges students, educators, and ChatGPT.
    \item \textbf{Knowledge Base:} This is a repository of relevant information and resources, including additional teaching materials, that students can access to enhance their learning. The repository also records prompts and responses between the students and ChatGPT and submitted tasks. The latter allows educators to compare the similarity and originality of students’ work if necessary.
    \item \textbf{Problem Solver:} Students may require step-by-step guidance to solve problems, e.g. to solve derivation in calculus or to compose a recursive algorithm. These step-by-step guidance should be prepared in advance by educators as additional teaching materials or as pre-prepared~prompts.
\end{enumerate}

\section{FUTURE PROSPECTS}

Further exploration of the potential of LLMs to enhance the quality of higher education across multiple dimensions is essential. While this research has primarily focused on utilizing LLMs as learning assistants in various courses, it is equally important to investigate their potential in fostering the development of students' soft skills. As the demand for essential soft skills grows in response to rapid advancements in industry automation, widespread technology adoption, and the accelerating pace of technological innovation, job seekers must increasingly master skills such as analytical thinking, creative problem-solving, resilience, adaptability, self-motivation, and lifelong learning, among others~\cite{dibattista2023future}.

While much of the current research on AI in education focuses on improving course content and teaching methodologies, future work should extend this focus to examine how LLMs can facilitate the acquisition of these critical soft skills. In particular, attention should be given to those skills that require collaboration and interpersonal interaction, areas where LLMs may offer unique support for students preparing for the dynamic and evolving workforce.


\bibliographystyle{IEEEtran}
\bibliography{template}

\begin{thebibliography}{10}
\providecommand{\url}[1]{#1}
\csname url@samestyle\endcsname
\providecommand{\newblock}{\relax}
\providecommand{\bibinfo}[2]{#2}
\providecommand{\BIBentrySTDinterwordspacing}{\spaceskip=0pt\relax}
\providecommand{\BIBentryALTinterwordstretchfactor}{4}
\providecommand{\BIBentryALTinterwordspacing}{\spaceskip=\fontdimen2\font plus
\BIBentryALTinterwordstretchfactor\fontdimen3\font minus \fontdimen4\font\relax}
\providecommand{\BIBforeignlanguage}[2]{{%
\expandafter\ifx\csname l@#1\endcsname\relax
\typeout{** WARNING: IEEEtran.bst: No hyphenation pattern has been}%
\typeout{** loaded for the language `#1'. Using the pattern for}%
\typeout{** the default language instead.}%
\else
\language=\csname l@#1\endcsname
\fi
#2}}
\providecommand{\BIBdecl}{\relax}
\BIBdecl

\bibitem{kowalski2020globalsouth}
\BIBentryALTinterwordspacing
A.~Kowalski, ``Global south-global north differences,'' in \emph{No Poverty}, ser. Encyclopedia of the UN Sustainable Development Goals, W.~Leal~Filho, A.~Azul, L.~Brandli, A.~Lange~Salvia, P.~Özuyar, and T.~Wall, Eds.\hskip 1em plus 0.5em minus 0.4em\relax Springer, Cham, 2020. [Online]. Available: \url{https://doi.org/10.1007/978-3-319-69625-6_68-1}
\BIBentrySTDinterwordspacing

\bibitem{uvalic2016sustainable}
S.~Uvalic-Trumbic and J.~Daniel, ``Sustainable development begins with education,'' 2016.

\bibitem{hesa2022world}
\BIBentryALTinterwordspacing
{Higher Education Strategy Associates (HESA)}, ``World higher education: Institutions, students and funding, main report,'' 2022. [Online]. Available: \url{https://higheredstrategy.com/world-higher-education-institutions-students-and-funding/}
\BIBentrySTDinterwordspacing

\bibitem{macgregor2022higher}
\BIBentryALTinterwordspacing
K.~MacGregor, ``Higher education report charts rise of the global south,'' \emph{University World News}, 2022, 12 March 2022. [Online]. Available: \url{https://www.universityworldnews.com/post.php?story=20220311151815827}
\BIBentrySTDinterwordspacing

\bibitem{buckner2021quantity}
E.~Buckner and Y.~Zhang, ``The quantity-quality tradeoff: a cross-national, longitudinal analysis of national student-faculty ratios in higher education,'' \emph{Higher Education}, vol.~82, pp. 39--60, 2021.

\bibitem{schenker2012economics}
A.~Schenker-Wicki and M.~Inauen, ``The economics of teaching: what lies behind student-faculty ratios?'' \emph{Higher Education Management and Policy}, vol.~23, no.~3, pp. 1--20, 2012.

\bibitem{minaee2024largelanguage}
S.~Minaee \emph{et~al.}, ``Large language models: A survey,'' \emph{arXiv preprint arXiv:2402.06196}, 2024.

\bibitem{liu2023harnessing}
B.~L. Liu \emph{et~al.}, ``Harnessing the era of artificial intelligence in higher education: a primer for higher education stakeholders,'' 2023.

\bibitem{kshetri2023future}
N.~Kshetri, ``The future of education: Generative artificial intelligence’s collaborative role with teachers,'' \emph{IT Professional}, vol.~25, no.~6, pp. 8--12, 2023.

\bibitem{li2024selfdirected}
B.~Li \emph{et~al.}, ``Reconceptualizing self-directed learning in the era of generative ai: An exploratory analysis of language learning,'' \emph{IEEE Transactions on Learning Technologies}, 2024.

\bibitem{lee2024chatgpt}
G.-G. Lee and X.~Zhai, ``Using chatgpt for science learning: A study on pre-service teachers' lesson planning,'' \emph{IEEE Transactions on Learning Technologies}, 2024.

\bibitem{chen2024multimedia}
X.~Chen and D.~Wu, ``Automatic generation of multimedia teaching materials based on generative ai: Taking tang poetry as an example,'' \emph{IEEE Transactions on Learning Technologies}, 2024.

\bibitem{rodriguez2024chatgpt}
R.~Rodriguez-Echeverr\'ia \emph{et~al.}, ``Analysis of chatgpt performance in computer engineering exams,'' \emph{IEEE Revista Iberoamericana de Tecnologias del Aprendizaje}, 2024.

\bibitem{wang2023unleashing}
M.~Wang \emph{et~al.}, ``Unleashing chatgpt's power: A case study on optimizing information retrieval in flipped classrooms via prompt engineering,'' \emph{IEEE Transactions on Learning Technologies}, 2023.

\bibitem{ahmad2023generative}
N.~Ahmad, S.~Murugesan, and N.~Kshetri, ``Generative artificial intelligence and the education sector,'' \emph{Computer}, vol.~56, no.~6, pp. 72--76, 2023.

\bibitem{mosaiyebzadeh2023chatgpt}
F.~Mosaiyebzadeh \emph{et~al.}, ``Exploring the role of chatgpt in education: applications and challenges,'' in \emph{Proceedings of the 24th Annual Conference on Information Technology Education}, 2023.

\bibitem{pintrich1988motivated}
P.~R. Pintrich and E.~V.~D. Groot, ``Motivated strategies for learning questionnaire,'' \emph{Journal of Educational Psychology}, 1988.

\bibitem{dibattista2023future}
\BIBentryALTinterwordspacing
A.~D. Battista, S.~Grayling, E.~Hasselaar, T.~Leopold, R.~Li, M.~Rayner, and S.~Zahidi, ``Future of jobs report 2023,'' 2023. [Online]. Available: \url{https://www.weforum.org/reports/the-future-of-jobs-report-2023}
\BIBentrySTDinterwordspacing

\end{thebibliography}

\end{document}